\documentclass{article}
\usepackage{xspace}
\usepackage[utf8]{inputenc} 
\usepackage{subfiles}
\usepackage{color}
\usepackage{amsmath}
\usepackage{amsfonts}
\usepackage{amssymb}
\usepackage{mathtools}
\usepackage{graphicx}
\usepackage{tikz}
\usepackage{pgfplots}
\usepackage[justification=centering,font=scriptsize,skip=0pt, belowskip=-1pt,aboveskip=0pt]{caption}
\usepackage[ruled, algosection, noend, linesnumbered]{algorithm2e}
\DeclarePairedDelimiter{\ceil}{\lceil}{\rceil}
\SetCommentSty{mycommfont}
\SetKwRepeat{Do}{do}{while}
\usepackage[noadjust]{cite}
\newcommand{\landmarkSet}{L}
\newcommand{\repDegree}{r}

\newcommand{\systemCapacity}{n\xspace}
\newcommand{\bw}{bw\xspace}
\newcommand{\mor}{\textit{Pyramid}\xspace}
\newcommand{\storage}{rpLoad\xspace}
\newcommand{\virtualSystemSize}{{vs}_{size}\xspace}
\newcommand{\timeSlot}{\textit{ts}\xspace}
\newcommand{\goal}{utility- and locality-awareness\xspace}

\newcommand{\fpti}{\textit{FPTI}\xspace}
\newcommand{\qostable}{\textit{UT}\xspace}

\newcommand{\availabilityGain}{1.2\xspace}
\newcommand{\localityGain}{1.1\xspace}
\usepackage{pifont}
\newcommand{\cmark}{\ding{51}}%
\newcommand{\xmark}{\ding{55}}%
\usepackage{amsmath}

\usepackage{amssymb}

\let\norm\undefined 
\DeclarePairedDelimiter\norm{\lVert}{\rVert}
\let\OLDthebibliography\thebibliography
\renewcommand\thebibliography[1]{
 \OLDthebibliography{#1}
 \setlength{\parskip}{0pt}
 \setlength{\itemsep}{0pt plus 0.3ex}
}

\begin{document}

\title{Decentralized utility- and locality-aware replication for heterogeneous DHT-based \\ P2P cloud storage systems}

\author{Yahya Hassanzadeh-Nazarabadi, Alptekin K\"{u}p\c{c}\"{u}, and \"{O}znur \"{O}zkasap\\Department of Computer Engineering, Ko\c{c} University, İstanbul, Turkey\\
{\{yhassanzadeh13, akupcu, oozkasap\}}@ku.edu.tr}

\maketitle
\begin{abstract}
As a Distributed Hash Table (DHT), Skip Graph routing overlays are exploited in several peer-to-peer (P2P) services, including P2P cloud storage. The fully decentralized replication algorithms that are applicable to the Skip Graph-based P2P cloud storage fail on improving the performance of the system with respect to both the availability of replicas as well as their response time. Additionally, they presume the system as homogeneous with respect to the nodes' latency distribution, availability behavior, bandwidth, or storage. In this paper, we propose \mor, which is the first fully decentralized utility- and locality-aware replication approach for Skip Graph-based P2P cloud storage systems. \mor considers the nodes as heterogeneous with respect to their latency distribution, availability behavior, bandwidth, and storage. \mor is utility-aware as it maximizes the average available bandwidth of replicas per time slot (e.g., per hour). Additionally, \mor is locality-aware as it minimizes the average latency between nodes and their closest replica. 
Our simulation results show that compared to the state-of-the-art solutions that either perform good in utility-awareness, or in locality-awareness, our proposed \mor improves both the utility- and locality-awareness of replicas with a gain of about \textbf{$\availabilityGain$} and \textbf{$\localityGain$} times at the same time, respectively.
\end{abstract}


\section{Introduction}
\label{mor:sec_introduction}
A peer-to-peer (P2P) cloud storage system consists of a set of peers, i.e., computing nodes that are interconnected over the Internet (e.g., computers, mobile devices, resource-constrained devices). There exist two roles in such cloud storage systems; data owner and data requester. A data owner holds a set of data objects and aims to share them with a group of authorized peers that are named data requesters. A data owner may also be a data requester for the other data owners. 

Distributed Hash Table (DHT)-based P2P cloud storage \cite{he2010study, hwang2009cloud, xu2009cloud} is a type of P2P cloud storage systems that operates over a structured overlay, where each peer is represented by a \textit{node} in the overlay. Each node knows a logarithmic number of other nodes (i.e., neighbors) in the system, and keeps them as $(ID, address)$ pairs in its \textit{lookup table}. Using its lookup table, each node is able to efficiently search and find other nodes in a fully decentralized manner. Skip Graph \cite{aspnes2007skip} is a DHT routing overlay that supports scalability, fast searching, and load balancing regarding message routing. Such features enable Skip Graph as a suitably structured overlay for DHT-based P2P cloud storage applications \cite{goyal2018adaptive, udoh2011cloud, batrashort, singh2015p, shabeera2012authenticated, hassanzadeh2016awake, hassanzadeh2015locality, hassanzadeh2016laras, hassanzadeh2018decentralized, boshrooyehguard}. Furthermore, Skip Graph can be considered as an alternative to other DHT overlays (e.g., Chord \cite{stoica2003chord}) in their DHT-based P2P services ranging from distributed storage systems \cite{rowstron2001pastry} to online social networks \cite{buchegger2009peerson} and search engines \cite{yang2006proof}. 

Nodes in P2P cloud storage systems are prone to churn, which is known as the dynamic arrival and departure behavior of the nodes \cite{stutzbach2006understanding}. Once a node arrives at the system, it is considered as available (online) until it departs the system (i.e., goes offline) or fails. A departed or failed node may join the system at a later time or may leave the system forever. Churn in the system results in data unavailability in P2P cloud storage \cite{rahmani2014comparative}, as upon departure or failure of a data owner, its data objects would no longer be available to the data requesters. In order to reduce the query load over the data owner, avoid the single point of failure, provide fault tolerance, and improve the data availability under churn, the data owner makes copies of its data objects on some other nodes of the system, which are called the data owner's replicas. The process of determining and managing the replicas is known as replication \cite{tanenbaum2017distributed}. Adapting a replication policy on the P2P cloud storage systems defines a collaborative environment where nodes participate with their idle storage spaces, and are in charge of storing each others' data objects in exchange of having their data objects stored on some other nodes of the system. 

The performance of P2P cloud storage is evaluated with respect to the accessibility of replicas \cite{shen2010efficient} and their availability under churn \cite{kermarrec2012availability}. The accessibility of replicas is correlated with their latency distribution in the underlying network, and is evaluated by the average access delay, which corresponds to the average latency between each data requester and its closest replica. The closest replica of each data requester is the one with the minimum Round-Trip-Time (RTT), and is called the \textit{corresponding replica} of that data requester. The replication approaches that aim at minimizing the average access delay of replicas are known as \textit{locality-aware} \cite{hassanzadeh2016laras}. 
The availability of replicas is correlated with their churn behavior, and is evaluated as their average availability per time slot, e.g., the average number of available replicas at each hour. The replication approaches that aim at maximizing the average availability of replicas are known as \textit{availability-aware} \cite{hassanzadeh2016awake}. Following these performance-oriented goals, the best P2P cloud storage is the one that provides both the locality- and availability-awareness of replicas at the same time. 

The locality- and availability-awareness of replicas have not been addressed together in a fully decentralized manner for DHT-based P2P cloud storage systems \cite{sun2017analysis, rahmani2014comparative}. By the full decentralization, we mean a strategy where the data owner is able to place its replicas without the need of any special node (e.g., supernodes) to handle the computation, storage, or communication that is needed for the replica placement. Our definition of full decentralization, hence, stands against the solutions such as \cite{rodolakis2006replicated, kalpakis2001optimal, tang2005qos, wu2017algorithms}, where a centralized entity is in charge of collecting the state of all nodes, executing the replication algorithm on behalf of the data owner, and deciding on the replica placement.

Considering the locality-awareness for Skip Graph-based P2P cloud storage systems, GLARAS \cite{hassanzadeh2018decentralized} is the best fully decentralized one. However, it does not support availability-awareness of replicas. Likewise, the traditional fully decentralized availability-aware replications that are applicable to Skip Graphs do not consider the average access delay of replicas. Such availability-aware replications are classified into reactive and proactive ones. The reactive replications improve the data availability by reactively resolving the departure or failure of replicas \cite{shen2007locality, datta2006internet, shen2010efficient, 10.1371/journal.pone.0205757, gopalakrishnan2004adaptive, paiva2015uto, paiva2011rollerchain}. While the proactive ones aim at providing an average number of available replicas for a long period of time, e.g., providing three available replicas on average for one month \cite{legtchenko2012relaxdht, su2004replica, ktari2007performance, harwood2003hashing, xiaosu2011caching, paiva2015data, pitoura2006replication, knevzevic2009dht, zaman2011distributed, pace2011exploiting, kermarrec2012availability, le2009finding}. 
It is worth mentioning that some of the existing availability-aware replications aim at minimizing the average number of intermediate nodes between replicas and data requesters \cite{shen2010efficient, pace2011exploiting}. However, as we experimentally demonstrated in \cite{hassanzadeh2015locality}, queries of the same path length show drastically different response time depending on the overlay connectivity. Hence, the number of intermediate nodes between the data requesters and replicas does not necessarily reflect the locality-awareness of replicas.


Evaluating the data availability of a P2P storage system by the availability-awareness of its replicas is only applicable to the homogeneous systems \cite{meroufel2013managing, wu2008optimal, aral2018decentralized, legtchenko2012relaxdht, kermarrec2012availability}, where nodes participate with similar bandwidths. In such systems, a larger number of online replicas reflects a higher available bandwidth, and hence a better level of data availability for the data requesters. In the real world P2P systems, where nodes participate with heterogeneous bandwidths, a larger number of online replicas does not necessarily imply a better level of data availability. For example, two nodes with the available bandwidth of $100Kbps$ are not good replication candidates for serving $100$ data requesters simultaneously. As each data requester only benefits from an average concurrent bandwidth of $2Kbps$. In the same example, a single node with an available bandwidth of $1Mbps$ provides an average bandwidth of $10Kbps$ for each data requester, which results in a better level of data availability. We conclude that addressing the data availability in the systems with heterogeneous bandwidth requires a stronger performance metric, which should consider both the availability of replicas as well as their bandwidth heterogeneity. To address this issue, we introduce the \textbf{utility-awareness} of replicas as a stronger performance metric (than their availability-awareness) for the bandwidth heterogeneous P2P cloud storage systems under churn. To define utility-awareness, we consider dividing a fixed-size periodic time interval (\textit{\fpti}) into a set of identical time slots (\textit{\timeSlot}), e.g., \textit{FPTI} as a day and \textit{\timeSlot} as an hour. We then evaluate the utility-awareness of replicas as the average available bandwidth of replicas during each \textit{\timeSlot} where the average is taken over the corresponding data requesters of each online replica during each \textit{\timeSlot}. Accordingly, we consider a replication mechanism as \textbf{utility-aware} if it aims at maximizing the utility-awareness of replicas. Similar to the availability-awareness, the utility-awareness is also correlated with the individual features of the nodes, i.e., availability behavior and bandwidth. 

In this paper, to improve the data availability and data accessibility of the heterogeneous DHT-based P2P cloud storage systems, \textbf{we propose \mor, which is the first fully decentralized utility- and locality-aware replication algorithm for Skip Graph-based P2P cloud storage systems}. \mor considers the nodes as heterogeneous with respect to their latency distribution in the underlying network, availability behavior, bandwidth, and storage capacity.
By employing \mor, a data owner can replicate its data objects in a fully decentralized manner with the maximized \goal of its replicas is achieved. Since Skip Graphs can be used as alternatives to other DHTs in their P2P services, by employing \mor, any DHT-based application can benefit from a utility- and locality-aware replication service, e.g., DHT-based cloud storage systems \cite{he2010study, hwang2009cloud, xu2009cloud, rowstron2001pastry, stoica2003chord},  search engines \cite{yang2006proof}, and social networks \cite{buchegger2009peerson}.

The original contributions of this paper are as follows: 
\begin{itemize}
    \item We propose \mor \footnotemark: the first fully decentralized, proactive, utility-, and locality-aware replication algorithm for heterogeneous Skip Graph-based P2P cloud storage systems. 
    \footnotetext{\mor is the extension of our earlier work, Awake (\textit{DOI: https://doi.org/10.1109/SmartCloud.2016.45}), which solely supports the availability awareness of replicas.} \mor provides an optimization model that aims on maximizing both the utility- and locality-awareness of replicas.
    \item We extended the Skip Graph simulator, SkipSim \cite{skipsim}, for simulating and evaluating the \goal of replication algorithms. 
    We implemented the best existing fully decentralized availability- and locality-aware replication algorithms on SkipSim, adapted them to our system model, and compared their performance against our proposed \mor.
    \item Based on our simulation results and compared to the best existing decentralized replications that either perform good in utility-awareness, or in locality-awareness, \mor improves the \goal of the replicas with a gain of about \textbf{$\availabilityGain$} and \textbf{$\localityGain$} at the same time, respectively. 
\end{itemize}

\section{Preliminaries}
\label{pyramid:sec_preliminaries}
\textbf{Skip Graph: }Skip Graph \cite{aspnes2007skip} is a DHT overlay where each node has two identifiers: a numerical ID and a name ID. Numerical IDs are non-negative integers, and name IDs are binary strings. The basic operations in a Skip Graph overlay are the search for numerical ID \cite{aspnes2007skip} and search for name ID \cite{hassanzadeh2016laras}, which enable a (search initiator) node to look for the owner of a specific numerical ID or name ID, respectively. Having $\systemCapacity$ nodes in a Skip Graph, a search initiator is able to perform both searches in a fully decentralized manner, and with the communication complexity of $O(\log{\systemCapacity})$. As the result of a search for a numerical ID, (IP) address of the node with the largest numerical ID that is less than or equal to the search target is returned to the search initiator. As the result of a search for name ID, (IP) address(es) of the node(s) with the longest common \textbf{prefix} with the search target are returned to the search initiator. We elaborate more on the search for name ID in Section \ref{pyramid:sec_pyramid}. In this paper, we define the \textit{system capacity} as the maximum number of the registered nodes to the Skip Graph, denoted by $\systemCapacity$.

\textbf{Locality-Aware Skip Graph: } In a landmark-based Skip Graph \cite{hassanzadeh2015locality} the overlay is virtually divided into a number of \textbf{regions}. Each region is recognized by a single landmark. \textbf{Landmarks} are not Skip Graph nodes, rather they are external components (e.g., servers) that are solely employed as reference points. Nodes use landmarks as some external reference points to ping, measure their RTT with respect to them, and share it with other nodes. Hence, using a landmark-based Skip Graph does not imply any sort of centralization at all, and the system is still administrated by the fully decentralized Skip Graph overlay of nodes. Studies like \cite{boshrooyeh2018distributed} also propose decentralized approaches to use the nodes themselves as the landmarks instead of relying on external ones. Each node of the Skip Graph belongs to the region of its closest landmark, which is the one with the minimum associated RTT. We denote the set of landmarks by $\landmarkSet$ and assume that $|\landmarkSet| = O(\log{\systemCapacity})$. We also assume that the set $\landmarkSet$ is constant over time, and is known by every node of the system.  In a landmark-based locality-aware Skip Graph overlay \cite{hassanzadeh2018decentralized}, the name IDs of the nodes are assigned in a way that the length of the common prefix in the name IDs of nodes is an inverse function of their RTT, i.e., a longer common prefix in name IDs of two nodes reflects their lower RTT in the underlying network. For example, in a locality-aware Skip Graph, a node with name ID of $0011$ is expected to experience a lower latency to the node with name ID of $0001$ than the node with name ID of $0111$. Since the name ID of $0011$ has a $2$-bit common prefix length of $00$ with $0001$, while it has only a $1$-bit common prefix length of $0$ with $0111$. 

\textbf{System-Wide Distribution of replicas (SWD): }
SWD is an independent module of the GLARAS replication algorithm \cite{hassanzadeh2018decentralized}, which approximates the optimal distribution of the replication degree among the regions of the system considering the locality-awareness of the replication. The replication degree denotes the number of replicas a data owner aims to place. A data owner executes SWD in a fully decentralized manner. Given the replication degree and the set of landmarks, SWD distributes the replication degree among the regions of the system based on some approximation on the data requesters' distribution in the Skip Graph overlay. As a result of SWD, each region of the system receives a sub-replication degree denoting the number of replicas should be placed in that region of the system.

\textbf{Churn Models: }
In P2P systems nodes are transient between online and offline states. The online and offline dynamics of the nodes is described by a churn model \cite{stutzbach2006understanding, jimenez2009connectivity, laredo2008resilience, guo2005measurements}. A churn model is specified by two distributions; session length, and inter-arrival time. The session length distribution characterizes the duration of the nodes' online states in the system. The inter-arrival time distribution characterizes the time interval between two consecutive arrivals of nodes to the system. Availability of a node is correlated with its session length, i.e., a longer average session length for a node corresponds its higher availability in the system. The availability of the system itself is correlated with the inter-arrival time distribution, i.e., a shorter average inter-arrival time corresponds a larger number of arrivals in the unit of time, which makes the system more available in the terms of the number of online nodes in the system.

\section{System Model and Overview}
\label{pyramid:sec_system_model}
\textbf{System Model:}
Upon arrival to the system, a peer represents itself as a Skip Graph node by assigning its numerical ID as the hash value of its (IP) address, and its name ID by invoking a locality-aware name ID assignment scheme (e.g., LANS \cite{hassanzadeh2018decentralized}), and joins the Skip Graph overlay using the join protocol in a fully decentralized manner \cite{aspnes2007skip}. Nodes use the Skip Graph overlay to discover each others' (IP) addresses by performing searches for each others' name IDs \cite{hassanzadeh2016laras} or numerical IDs \cite{aspnes2007skip}. 
We assume that the nodes are heterogeneous with respect to their storage capacity, bandwidth, availability behavior, and latency distribution in the underlying network. Each online node frequently computes its utility state, which is a function of its available storage capacity, bandwidth, and availability behavior, and shares it with the other nodes using an aggregation mechanism. 
The aggregation scheme acts as a shared fully decentralized bulletin board, which keeps the aggregated utility state of the system. 

In our system model, we assume that nodes depart the system arbitrarily by invoking the departure protocol of the Skip Graph in a fully decentralized manner \cite{aspnes2007skip}. The departure protocol keeps the connectivity of the overlay under churn. However, unnoticed departures can be handled via a decentralized churn stabilization algorithm (e.g., Interlaced \cite{hassanzadeh2019interlaced}). Also, we assume the storage capacity of nodes is discrete in the unit of storage, and a data owner aims at utilization of one storage unit of a node for replication of its data object. Following this assumption, for example, if a node owns a free storage capacity of $3$ units, it can be the replica of at most $3$ data owners. 
Besides, we assume that all nodes are the data requesters of every data owner. Note that managing the access control of replicas is an orthogonal issue that, for example, can be handled by the data owner encrypting the data objects and sharing the key with the (authorized) data requesters.


\textbf{Utility Vector:} 
We model the utility state of a node at each time slot as a function of its available storage, bandwidth, and availability probability. We represent the utility state of the node $i$ by the vector $UV_{i}$, named its \textit{utility vector}. Size of the utility vector corresponds to the number of time slots of length $\timeSlot$ during one cycle of $\fpti$ i.e., $|UV_{i}| = \frac{\fpti}{\timeSlot}$. For example, considering $\fpti$ as a day and $\timeSlot$ as an hour, utility vector of a node is of size $24$. $UV_{i,t}$ represents the utility of the node $i$ during the $t^{th}$ time slot of the $\fpti$ (e.g., $t^{th}$ hour of the day), and is computed as shown by Equation \ref{pyramid:eq_av}. In this equation, $p_{i,t}$ is the availability probability of node $i$ during the $t^{th}$ time slot of \fpti, $bw_{i}$ is its normalized bandwidth, and $\storage_{i}$ is its replication load.

\begin{equation}
    UV_{i,t} = \frac{p_{i,t} \times \bw_{i}}{\storage_{i} + 1}
    \label{pyramid:eq_av}
\end{equation}

Node $i$ computes $p_{i,t}$ by dividing the overall time it has been online during the $t^{th}$ time slot over the number of times that $\fpti$ has cycled up to the computation time. For example, considering $\fpti$ as a day and $\timeSlot$ as an hour where $7$ days have elapsed (i.e., $\fpti$ has cycled $7$ times), $p_{i,t}$ denotes the overall fraction of time that node $i$ was online at the $t^{th}$ hour of the day, over the past $7$ days. If the node $i$ was online for $3$ hours during the $t^{th}$ hour in the past $7$ days, $p_{i,t}= \frac{3}{7} = 0.42$. 

We define the \textit{replication load} of node $i$ (i.e., $\storage_{i}$) as the number of data owners that it has already been designated as their replica. The purpose of defining the replication load is two-fold. First, the replication load determines the amount of storage that a node devotes to serve as the replica. Second, one can compare the strength of the bandwidth that two replicas provide per data requester on the average by comparing their corresponding replication loads. For example, for two replica nodes with an identical bandwidth, a data requester is likely to get a better bandwidth from the one with lower replication load, as the loosely loaded replica likely has to respond to less number of data requesters' queries compared to the heavily loaded one.  

By the normalized bandwidth, we mean the bandwidth of a node is mapped to a value in $[0,1]$ by dividing that over the maximum bandwidth in the system. The maximum bandwidth is considered as a system-wide constant parameter of the protocol. 
The $+1$ in the denominator of Equation \ref{pyramid:eq_av} is to prevent division by zero when node $i$ has not been selected as a replica yet (i.e., $\storage_{i} = 0$). $UV_{i,t}$, hence, is taking a value in $[0,1]$. The higher the $UV_{i,t}$ is, the higher is the utility node $i$ provides in the replication process.


\textbf{Replication with \mor:} 
Having the aggregated utility state of the system, each data owner is able to invoke the \mor algorithm when it needs replication for its data. As the result, \mor determines the (IP) address of replicas in a fully decentralized manner considering the \goal of the replication. 
The data owner then shares its list of replicas on the decentralized aggregation scheme, which makes them publicly available for its data requesters.  
Inputs to \mor are the utility states of the nodes in the system, the replication degree of the data owner, and the information about the landmarks. \mor first distributes the replication degree among the regions of the system, where each region receives a sub-replication degree. For each region, \mor then models the \goal of replication with Integer Linear Programming (ILP), solves it and finds the placement of replicas, accordingly. 
The challenge in designing \mor is that obtaining, storing, and operating on the utility vectors of all the nodes result in an asymptotic linear dependency of the communication, storage, and time complexities on the system capacity. Considering the data owner that executes \mor as a resource-constrained peer, these linear dependencies would cause performance degradation.

To improve the communication complexity, \mor is built upon the existence of an aggregation mechanism with  $O(\log{\systemCapacity})$ communication complexity. The aggregation scheme acts as a decentralized bulletin board, and enables each node to write its utility vector on it as well as to retrieve all the utility vectors of other nodes from it with $O(\log{\systemCapacity})$ communication complexity. In this paper, we employ LightChain \cite{hassanzadeh2019lightchain}, as the underlying aggregation scheme. LightChain is a churn resilient DHT-based blockchain with $O(\log{\systemCapacity})$ communication complexities on storing and retrieving the latest state a data object. Besides, LightChain distributes the storage of the utility vectors uniformly among the nodes, so no node is required to keep the entire blockchain database. 

To improve the time and storage complexities, \mor squeezes the \textit{original system} of size $\systemCapacity$ nodes to a significantly smaller size system of size $\log{\systemCapacity}$ nodes that is called the \textbf{virtual system}. \mor provides an efficient many-to-one mapping functionality for each original node to find its corresponding virtual node. The utility vector of a virtual node represents the average of the utility vectors of all its corresponding original nodes. The set of utility vectors of all the virtual nodes is stored on the aggregation scheme and gets updated frequently by the original nodes. We call this set the \textit{aggregated utility state of the system}. Each original node is able to read the latest aggregated utility state of the system as well as to update it, i.e., by integrating its own latest utility vector to the utility vector of its corresponding virtual node. Instead of operating on the individual utility vectors of the original nodes, which requires linear storage and time complexity in the system size, \mor receives the aggregated utility state of the system and operates solely on it. As we discuss in Section \ref{pyramid:sec_results}, mapping to the virtual system and operating on it results in the storage and time complexities of $O(\log{\systemCapacity})$ and $O(\log^{2}{\systemCapacity})$ for \mor, respectively.


\section{Details of \mor}
\label{pyramid:sec_pyramid}
\subsection{Virtual System}
As explained in Section \ref{pyramid:sec_system_model}, to perform the computation efficiently, \mor does not directly operate on the original system. Rather, it operates on the virtual system, which is the squeezed model of the original system. The virtual system has the same set of landmarks and regions as the original system. Except, multiple original nodes (i.e., nodes in the original system) are mapped into a single node of the virtual system. The size of the virtual system is denoted by $\virtualSystemSize$ and is assumed as a protocol parameter known by all the nodes. As we discuss later, we consider $\virtualSystemSize = \log{\systemCapacity}$. The virtual nodes (i.e., nodes in the virtual system) have an identifier length of $\ceil{\log{\virtualSystemSize}}$ bits. All the original nodes that have $\ceil{\log{\virtualSystemSize}}$ bits common prefix in their name IDs are represented by a virtual node. For example, assuming that $\virtualSystemSize = 16$, virtual nodes have $4$-bits identifiers, and all the original nodes with a name ID prefix of $0001$ are mapped to the virtual node $0001$. So, each original node is able to identify its corresponding virtual node by only taking the first $\ceil{\log{\virtualSystemSize}}$ bits of its own name ID. We denote the associated virtual node of the original node $j$ by \textit{virtual($j$)}.

\subsection{Mapping to the virtual system}
Mapping from the original system to the virtual system is done by the original nodes computing a utility vector for their corresponding virtual nodes in a fully decentralized manner by means of the underlying aggregation scheme of the system. The set of the utility vectors of the virtual nodes corresponds to the aggregated utility state of the system and is denoted by the table $\qostable$, which is named the \textit{utility table of the system}. $\qostable$ is a 3-dimensional table of size $|\landmarkSet| \times \virtualSystemSize \times |UV|$. $\qostable_{l,i,t}$ represents the utility vector of the virtual node $i$ in the region $l$ of the virtual system within the $t^{th}$ time slot of \fpti. The utility vector of the virtual node $i$ corresponds to the average of the utility vectors of all the original nodes that are mapped to it. $\qostable$ keeps the utility vector of virtual node $i$ as the number of its corresponding original nodes as well as the summation of their utility vectors. In this way, while the average utility vector corresponding to each virtual node is efficiently computable, the number of corresponding original nodes to each virtual node is also known. As we explain later, \textit{\mor} employs such information on replica placement. $\qostable$ is shared among the original nodes and maintained in a fully decentralized manner using the underlying aggregation scheme (i.e., LightChain blockchain \cite{hassanzadeh2019lightchain}). $\qostable$ is reset to zero at the beginning of each cycle of $\fpti$. Each online original node $j$ that belongs to the region $l$ of the original system and has free storage space, updates $\qostable$ once during each cycle of $\fpti$. The update is done by including the updated utility vector of the original node $j$ in the average utility vector of its corresponding virtual node, i.e.,  $\qostable_{l, virtual(j)}$. The underlying aggregation scheme also enables each node to efficiently retrieve the latest state of $\qostable$ on demand. 


\begin{algorithm}
{
\KwIn{Set of landmarks $\landmarkSet$, replication degree $\repDegree$, utility table of system $\qostable$}
\KwOut{Set of replicas in orginal system $oRepSet$}
\BlankLine
    \tcp{System-wide distribution of replicas}
    $R$   = SWD($\landmarkSet$, $\repDegree$)\;
    \label{pyramid_alg:pyramid_swd}

    \For{each region $l \in \landmarkSet$}
    {
        \tcp{Computing the time slot coverage weights}
        $W_{l} =$ TCWD$(\qostable_{l})$\; 
        \label{pyramid_alg:pyramid_tcwd}
        \tcp{Replicas placement in region $l$ of virtual system}
        $vRepSet = vRepSet \quad \cup$ RWD$(\qostable_{l}\,, R_{l}\,, W_{l})$\;
        \label{pyramid_alg:pyramid_ILP}
    }
    \For{each virtual replica $vRep \in vRepSet$}
    {
        \tcp{Finding best original node for virtual replica $vRep$}
        $oRep = searchForUtility(vRep)$\; 
        \label{pyramid_alg:pyramid_search}
        add $oRep$ to $oRepSet$\;
        \label{pyramid_alg:pyramid_bestmatch}
    }
    publish $oRepSet$ on the aggregation scheme\;
    \label{pyramid_alg:pyramid_aggregation}
\Indp
\Indm
\caption{\mor}
\label{pyramid:alg_Pyramid}
}
\end{algorithm}

\subsection{\mor Algorithm}
\textbf{Inputs and outputs:} \mor is represented by Algorithm \ref{pyramid:alg_Pyramid}, and is executed by a data owner to determine its replicas. The inputs to \mor are the set of landmarks' features (i.e., $\landmarkSet$), the replication degree (i.e., $\repDegree$), and the utility table of the system (i.e., $\qostable$). By the landmarks' features, we mean their (IP) addresses and pairwise latencies, which are assumed as public static information of the system. As the output, \mor returns $oRepSet$, which is the set of identifiers of replicas in the original system.

\textbf{System-Wide Distribution of replicas (SWD) (Algorithm \ref{pyramid:alg_Pyramid}, Line \ref{pyramid_alg:pyramid_swd}):}
On receiving the inputs, \mor distributes the replication degree among the regions of the system using SWD \cite{hassanzadeh2018decentralized}. Given the set of landmarks and the replication degree, SWD returns the set of sub-replication degrees, denoted by $R$ where $R_{l}$ denotes the number of replicas that should be placed in the region $l$ of the system considering only the locality-awareness of replicas. We skip the details of SWD for sake of space, as the details of SWD are not required for understanding our proposed \textit{\mor}. The interested readers are referred to \cite{hassanzadeh2018decentralized} for details about SWD. 

\textbf{Time slots Coverage Weight Distribution (TCWD) (Algorithm \ref{pyramid:alg_Pyramid}, Line \ref{pyramid_alg:pyramid_tcwd}):} The utility of the virtual nodes may not be distributed uniformly among all the time slots. Some time slots may be covered with the majority of high-utility virtual nodes, while the rest may be covered by only a few high-utility ones, or even left uncovered. We call such time-slots that are covered with only a few nodes or even no node as \textit{poorly covered time slots}. To maximize the utility of replicas during the poorly covered time slots, \mor identifies the poorly covered time slots of each region by assigning a weight to each of its time slots. The weight assignment is done by invoking the TCWD function of \mor. Invoking TCWD on the utility table of region $l$ (i.e., $\qostable_{l}$) results in obtaining the vector $W_{l}$ of the utility coverage weights for that region. $W_{l,t}$ is a number in $[0,1]$ and represents the utility coverage weight of the $t^{th}$ time slot in region $l$ of the system, and is computed by Equation \ref{pyramid:eq_tcwd}. In this equation, $count_{l,t}$ is the number of virtual nodes of region $l$ that their utility value at the $t^{th}$ time slot is less than or equal to the average utility value. The average is taken over the entire utility table of region $l$. A larger $W_{l,t}$ denotes a lower utility coverage of the $t^{th}$ time slot of region $l$.  

\begin{equation}
    W_{l,t} = \frac{count_{l,t}}{\sum_{t=1}^{|UV|}count_{l,t}}
    \label{pyramid:eq_tcwd}
\end{equation}


\textbf{Region-Wide Distribution of replicas (RWD) (Algorithm \ref{pyramid:alg_Pyramid}, Line \ref{pyramid_alg:pyramid_ILP}):}
Once the time slot utility coverage weights for the region $l$ of the virtual system are determined, \mor invokes the RWD function. RWD determines the placement of replicas in region $l$ of the virtual system considering the \goal of replicas. To place the replicas, RWD constructs an ILP model that is represented by Equations \ref{pyramid:objective}-\ref{pyramid:cons5}. The presented ILP model is for region $l$ of the virtual system, and hence $l$ is constant. The only decision variables in this ILP are $X$ and $Y$, and the rest are constant scalars. The decision variable $X$ is a three-dimensional binary table of size $\virtualSystemSize \times \virtualSystemSize \times |UV|$. $X_{i,j,t} = 1$ if the ILP assigns the virtual node $i$ as the corresponding replica for the virtual node $j$ during the $t^{th}$ time slot of $\fpti$, otherwise, $X_{i,j,t} = 0$. $Y$ is a one-dimensional decision variable of size $\virtualSystemSize$. $Y_{i} = 1$ if the ILP decides to place a replica on the virtual node $i$, otherwise $Y_{i} = 0$. 
$C$ is a three-dimensional table of the same size as $X$ that is constructed by the RWD function. $C_{i,j}$ represents the common prefix length between the name IDs of the virtual nodes $i$ and $j$. Name ID of the virtual node $i$ corresponds to the binary representation of $i$ in $\ceil{\log{\virtualSystemSize}}$ bits. Since \mor operates on a locality-aware overlay, $C_{i,j}$ conveys an approximated inverse quantification of the latency between the nodes $i$ and $j$ in the virtual system, i.e., a higher $C_{i,j}$ value implies a lower latency. 

\begin{align}
\max \sum_{t = 1}^{|UV|} \sum_{i = 1}^{\virtualSystemSize} \sum_{j = 1}^{\virtualSystemSize}&  X_{i,j,t} C_{i,j} \qostable_{l,i,t} W_{l,t} \quad \textrm{s.t.} \label{pyramid:objective} \\
\forall t \in [1, |UV|], i, j \in [1, \virtualSystemSize] \quad &Y_{i} \geq X_{i,j,t}  \label{pyramid:cons1}\\
\forall t \in [1, |UV|], i \in [1, \virtualSystemSize] \quad &\sum_{t = 1}^{|UV|} \sum_{j = 1} ^{\virtualSystemSize} X_{i,j,t} \geq Y_{i} \label{pyramid:cons2}\\ 
\forall t \in [1, |UV|], j \in [1, \virtualSystemSize] \quad &\sum_{i = 1} ^{\virtualSystemSize} X_{i,j,t}  = 1 \label{pyramid:cons3}\\
&\sum_{i = 1}^{\virtualSystemSize} Y_{i} = R_{l} \label{pyramid:cons4}\\
\forall t \in [1, |UV|], i, j \in [1, \virtualSystemSize] \quad &Y_{i} \in\{0,1\}, \quad X_{i,j,t}\in\{0,1\} \label{pyramid:cons5}
\end{align}

Equation \ref{pyramid:objective} shows the objective function of the \mor's ILP that aims at maximizing both the \goal of replicas. By maximizing the locality-awareness, we mean maximizing the summation of the name IDs' common prefix length between the data requesters and their corresponding replicas, which corresponds to minimizing the overall access delay of replicas. In Equation \ref{pyramid:objective}, when $X_{i,j,t} = 1$, virtual node $j$ experiences an access delay inversely proportional to $C_{i,j}$, while benefits from the (maximum) utilization proportional to $\qostable_{l,i,t}$. We consider the contribution of this replica assignment (i.e., virtual node $i$ as the corresponding replica of the virtual node $j$) to both \goal of the replication proportional to $C_{i,j} \times \qostable_{l,i,t}$ at every time slot $t$. 
To lead the ILP on maximizing the utility of the selected replicas at each time slot $t$, in the objective function, we project the utility of each virtual node with the associated utility coverage weight in that time slot, i.e., $W_{l,t}$. In this way, the ILP prioritizes to select the high-utility virtual nodes that also cover the poorly covered time slots, and improves the utility-awareness of replicas per time slot. 

Equations \ref{pyramid:cons1} and \ref{pyramid:cons2} represent the replica assignment constraints of the \mor's ILP. Equation \ref{pyramid:cons1} implies that a virtual node $j$ should be assigned to the virtual node $i$ as its corresponding replica during the $t^{th}$ time slot (i.e., $X_{i,j,t} = 1$) only if node $i$ itself is selected as a replica (i.e., $Y_{i} = 1$). 
Equation \ref{pyramid:cons2} says that if virtual node $i$ is selected as a replica (i.e., $Y_{i} = 1$), then it should be the corresponding replica of at least one other virtual node during at least one time slot. 

Equation \ref{pyramid:cons3} represents the data requester constraint of the \mor's ILP. It implies that at each time slot $t$, each node should be assigned to exactly one replica. This is done to lead the ILP towards finding the best replica that maximizes the objective function for the node. At each time slot $t$ and for each virtual node $j$ the summation of $X_{i,j,t}$ values over all $i$ values determines the number replicas that node $j$ is benefiting from during that time slot, which should be exactly one replica considering this constraint. 

Equation \ref{pyramid:cons4} shows the constraint on the sub-replication degree of region $l$. 
The summation in this equation corresponds to the overall number of permissible replicas that the data owner grants \mor to place in the region $l$ of the system, which should be exactly equal to the sub-replication degree of the region, i.e., $R_{l}$.  

Equation \ref{pyramid:cons5} shows the constraint on the legitimate values of the decision variables. That is, the only values that elements of $X$ and $Y$ decision variables can take are either $0$ or $1$. 

RWD solves the described ILP model and determines the placement of replicas in each region $l$ of the virtual system. The identifiers of selected virtual nodes by RWD as replicas are collected into the $vRepSet$, which contains the union of the identifiers of replicas in all regions of the virtual system. 

\textbf{Virtual to original system mapping of replicas (Algorithm \ref{pyramid:alg_Pyramid}, Lines \ref{pyramid_alg:pyramid_search}-\ref{pyramid_alg:pyramid_aggregation}):}
Each virtual replica identifier in $vRepSet$ corresponds to the name ID prefix of several nodes in the original system. Although replicating on any of these nodes satisfies the locality-awareness goal of \mor, not all of them may be suitable considering the utility-awareness goal. This follows the fact that the name IDs are assigned based on the locality information of the nodes, and do not reflect any utility attribute of them. Additionally, contacting all the original nodes with a name ID corresponding to a prefix and picking the one with the best utility requires linear time and communication complexities, and is not a scalable solution. To efficiently map a virtual replica to an original node with desirable utility, we develop a modified version of the search for name ID protocol \cite{hassanzadeh2016laras}, which we call the \textit{search for utility}. Given a virtual replica identifier $vRep \in vRepSet$, our search for utility protocol traverses at most $\alpha$ original nodes with the name ID prefix of $vRep$, and returns the one with the maximum utility score as the corresponding original node to $vRep$. The utility score of each original node $i$ corresponds to the second norm of its utility vector (i.e., $\norm{UV_{i}}$). We assume $\alpha$ is a system-wide constant parameter of the search for utility protocol. We denote the output of search for utility by $oRep$, which is the corresponding original node of the virtual replica $vRep$. On receiving $oRep$ from search for utility, \textit{\mor} adds it to the set of replicas in the original system, i.e., $oRepSet$. 




Once the mapping to the original system is done for all the virtual replicas, \mor publishes the $oRepSet$ on the underlying aggregation scheme. This enables the data requesters to query the aggregation scheme with the identifier of the data owner and retrieve its set of replicas. 
\section{Related Works}
\label{pyramid:sec_relatedworks}
\begin{table*}
\centering
\resizebox{\columnwidth}{!}
{
    \begin{tabular}{ |l|c|c|c|  }
    \hline
    Strategy & Availability-Awareness & Locality-Awareness & Utility-Awareness \\ 
    \hline
    Replication on path \cite{shen2010efficient, 10.1371/journal.pone.0205757, gopalakrishnan2004adaptive, paiva2015uto} & Reactive & \xmark & \cmark\\
    Virtual nodes \cite{paiva2011rollerchain} & Reactive & \xmark & \cmark\\
    Replication on neighbors \cite{shen2007locality,aral2018decentralized} & Reactive & \xmark & \cmark \\
    Randomized \cite{legtchenko2012relaxdht, su2004replica, ktari2007performance, harwood2003hashing, xiaosu2011caching, paiva2015data, pitoura2006replication, knevzevic2009dht} & Proactive & \xmark & \xmark\\
    Power-of-choice \cite{simon2014splad, simon2015scattering} & Proactive & \xmark & \cmark\\
    Cluster-based \cite{zaman2011distributed, pace2011exploiting} & Proactive & \xmark & \cmark \\
    Correlation-based \cite{kermarrec2012availability, le2009finding} & Proactive & \xmark & \cmark \\ 
    GLARAS \cite{hassanzadeh2018decentralized} & -- & \cmark & \xmark \\
    \textbf{\mor} & \textbf{Proactive} & \textbf{\cmark} & \textbf{\cmark}\\
    \hline
    \end{tabular}
}
\caption{Comparison of various decentralized replication algorithms}
\label{pyramid:table_compare}
\end{table*}

\subsection{Decentralized Availability-Aware Replication}
\textbf{Reactive replication} first performs an initial placement of replicas, and then the replicas are frequently probed to detect the failure events. A replica that is not answering the probe message in a certain while is presumed as failed, and is substituted by a newly placed one \cite{ktari2007performance}. Replication on path, virtual nodes, and replication on neighbors are the well-known reactive replication mechanisms. In replication on path \cite{shen2010efficient, 10.1371/journal.pone.0205757, gopalakrishnan2004adaptive, paiva2015uto}, nodes piggyback their query load on the search messages they route or initiate. The replication on search path then reacts to the failure of replicas by creating a new replica and relocates the existing replicas upon detection of a new query traffic hub. In the virtual nodes approach \cite{paiva2011rollerchain}, several peers are mapped to a single virtual node on the DHT overlay. The corresponding peers of the same virtual node all keep the same set of data objects. Virtual nodes approach reacts to the replication load of the peers by frequently monitoring their loads, splitting the heavily loaded groups into two, and re-distributing the data objects among them. In replication on neighbors \cite{shen2007locality,aral2018decentralized}, each node frequently checks its neighbors to find better ones in terms of replication load and availability, and relocates its replicas to its more available and loosely loaded neighbors. Relocating replicas upon failure is the common drawback of the reactive replication \cite{rahmani2014comparative}. Each reactive relocation applies a communication overhead to the system that is linear in the size of the replicated files.

\textbf{Proactive replication} aims at providing an average availability of replicas for a long period of time, for example, a few months. At the end of the scheduled period, a proactive mechanism should re-evaluate the placement of replicas based on the new state of the system and relocate the replicas if needed. Hence, proactive replication also does replica relocation but at a drastically slower pace. 
Power-of-choice is an enhanced version of randomized replication where the data owner randomly selects two replica candidates and replicates on the least loaded one \cite{simon2014splad, simon2015scattering}. Cluster-based replication \cite{zaman2011distributed, pace2011exploiting} groups the nodes based on features like availability pattern, replication load, and bandwidth, and distributes the replication degree among the clusters. In correlation-based replication \cite{kermarrec2012availability, le2009finding}, the replication is done on the pairs of anti-correlated nodes, i.e., pairs of nodes where the unavailability of one implies the availability of the other one with a high probability.


\subsection{Decentralized Locality-Aware Replication}
In \cite{hassanzadeh2018decentralized} GLARAS is proposed and experimentally shown that it is the best in locality-awareness compared with the existing decentralized locality-aware replication algorithms that are applicable on Skip Graphs. Although GLARAS is scalable with the ILP size of $O(\log{\systemCapacity})$, it does not consider the availability and utility of the nodes in replication. We skip surveying the other existing locality-aware replication algorithms for the sake of space and refer the interested readers to \cite{hassanzadeh2016laras, hassanzadeh2018decentralized} for a detailed discussion of locality-aware replication schemes.

Table \ref{pyramid:table_compare} summarizes a comparison between the existing decentralized replication algorithms and our proposed \mor. In this table, we mark an algorithm as utility-aware if it considers the bandwidth of the nodes in replica placement decision making. Compared to the existing solutions, \mor is the only proactive and fully decentralized replication algorithm for Skip Graph DHTs that provides \goal of replicas simultaneously.

\section{Simulation Setup}
\label{pyramid:sec_simulation}
We extended the Skip Graph simulator, SkipSim \cite{skipsim} by adding the aggregation functionality. We also enabled SkipSim to consider both the utility and locality of the nodes in scenarios with heterogeneous bandwidth and storage capacity of the nodes. 

\textbf{Churn model: }Among the existing churn models, the Bittorrent-based models in \cite{stutzbach2006understanding} are well-studied and parametrically clearer than the rest, e.g., \cite{datta2006internet, bustamante2004friendships, wu2008stochastic}. We hence implemented their churn model in SkipSim with Weibull distributions \cite{rinne2008weibull} for the session length and the inter-arrival time. In our implementation, all nodes follow the same Weibull-based session length distribution. However, to simulate different cliques of nodes with respect to the availability behavior, nodes of different regions follow distinct Weibull-based inter-arrival time distributions. Following our definition of regions in Section \ref{pyramid:sec_preliminaries}, this implies that nodes that are in the proximity of each other in the underlying network show a similar churn behavior. 
In our implemented churn model, considering the entire system, on average each node has a $2.7$ hours of session length followed by an offline period of $2.8$ hours.

\textbf{Bandwidth and Storage Capacity:} In our implementation, the bandwidth of the nodes follows an exponential distribution with an average of $2Mbps$. We extracted the shape of distribution from \cite{pouwelse2004measurement}, and adapted its scale with the typical average bandwidth provided by the service providers for the household customers \cite{kurose2016computer}. The storage capacity of the nodes is drawn from a uniform distribution in the range of $[1,3]$. We found this range as the one that quickly makes nodes out-of-storage by being selected as replicas of multiple data owners under randomized replication of $10$ data owners with replication degree of $14$. Hence, we establish our simulation setups on this range as a challenging range to decide on the replica placement for all the algorithms of interest. 
In our simulation, the storage capacity of a node denotes the maximum number of replicas that it can take from different data owners, e.g., a storage capacity of $2$ means that the node can be the replica of two data owners simultaneously. Once a node reaches its storage capacity, it simply stops sharing its utility vector as well as taking any further replication duty.

\textbf{Topologies: }We generated $5$ random topologies where each topology corresponds to a distinct distribution of the nodes in the Skip Graph overlay as well as their overlay connectivity. Each topology consists of $4096$ nodes, and $10$ randomly chosen data owners. In SkipSim, one simulation step corresponds to an hour. $\fpti$ and $\timeSlot$ were set to a day and an hour, respectively. Each topology was simulated for a lifetime of $3$ months. We spot a one-week learning phase for the data owners to learn the underlying utility behavior of the nodes. As we discuss in the followings the learning is done via aggregation and piggybacking. The data owners replicate at the end of this one-week learning period. 

\textbf{Replication algorithms: }We implemented our proposed \mor in SkipSim. Additionally, for the sake of comparison, we also implemented GLARAS, randomized replication, power-of-choice, cluster-based replication, and correlation-based replication with the implementation details provided in Section \ref{pyramid:sec_relatedworks}. For our proposed \mor, SkipSim performs the aggregation of utility vectors every $24$ time slots (i.e., hours). 
For the other algorithms, SkipSim provides $O(\systemCapacity_{o}^{2})$ random searches per time slot, where $\systemCapacity_{o}$ corresponds to the number of online nodes during that time slot. These random searches enable data owners to benefit from the piggybacked utility information of the nodes. By random searches, we mean the search initiator and the search target are chosen randomly among the set of online nodes. Each node on a search path piggybacks its utility vector to the search message, while it receives the updated piggybacked utility vectors of the previous nodes on the same path. 

\textbf{Virtual System Size:}
Based on our simulations for different values of the virtual system size in different system capacities, we found $\virtualSystemSize = 1.33\log{\systemCapacity}$ as a proper trade-off between the running time efficiency and performance of \mor in providing \goal of replicas, which is consistent with our previous work \cite{hassanzadeh2018decentralized}. Hence, during the simulations for the system capacity of $4096$ nodes, we set $\virtualSystemSize = 16$. Also, we simulated for different values of $\alpha$ in the search for utility procedure and found $\alpha = 3$ as a proper trade-off between the communication overhead of \mor and its performance in providing \goal of replicas in system capacity of $4096$ nodes. 
We skip the details for the sake of the page limit.

\subsection{Algorithms used for comparison}
For the sake of comparison with \mor, we choose 
the existing decentralized and proactive replications that are applicable to our system model, i.e., they do not change the operational complexity of the Skip Graph as well as its topology, do not require a centralized party, and are computationally efficient for a data owner peer to execute independently. For these algorithms, we implement piggybacking to enable each node disseminating its utility vector across the system efficiently. Also, as detailed in the followings, for all these algorithms except the randomized replication, we use the same utility scoring as our proposed \textit{\mor}. Each node in the system piggybacks its utility vector and (IP) address on the search messages it routes or initiates so that the receivers of the search messages can benefit from its utility vector. It is worth to mention that \textit{\mor} does not require the piggybacking and solely operates on the aggregated utility table of the nodes that is provided by the underlying aggregation scheme (i.e., LightChain blockchain \cite{hassanzadeh2019lightchain}). We discuss the asymptotic complexity of interacting with LightChain later in this paper. We also choose GLARAS as the best existing fully decentralized locality-aware replication for Skip Graph, and implement it exactly as specified in \cite{hassanzadeh2018decentralized}. Followings are the implementation details of the proactive replication algorithms.

\textbf{Randomized Replication \cite{legtchenko2012relaxdht}:}
The data owner chooses its replicas uniformly from the set of obtained addresses, pings each, and replicates on the online one that has free storage capacity. The data owner does this procedure repeatedly until it satisfies the replication degree.  

\textbf{Power-of-choice \cite{simon2014splad, simon2015scattering}:}
Our implementation of power-of-choice is similar to that of the randomized replication, except that to place a replica, the data owner chooses two nodes randomly, and replicates on the one that has the higher second norm of the utility vector. 

\textbf{Cluster-Based Replication \cite{zaman2011distributed, pace2011exploiting}:}
In our implementation of the cluster-based replication, the data owner performs an $\repDegree$-mean clustering \cite{kanungo2002efficient} of the nodes based on their piggybacked utility vectors, where $\repDegree$ is the replication degree of the data owner. The data owner then replicates on the nodes with the highest second norm of the utility vector from each cluster.

\textbf{Correlation-Based Replication \cite{kermarrec2012availability, le2009finding}:}
In our implementation of the correlation-based replication, the data owner aims to find $\frac{\repDegree}{2}$ pairs of nodes with the minimum correlation. To do this, the data owner first finds $\frac{\repDegree}{2}$ nodes with the maximum value of the second norm of the utility vector from its set of piggybacked utility vectors of nodes. For each of those nodes and in the same set of piggybacked utility vectors, the data owner finds the node with the highest fraction of the second norm of the utility vector over the correlation value, and replicates on both nodes. We measure the correlation between two nodes as the dot product of their utility vectors. The higher the dot product is, the higher the two nodes are correlated with each other. 
\section{Performance Results}
\label{pyramid:sec_results}


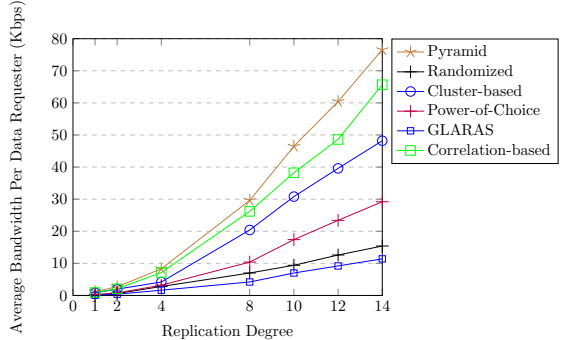
\begin{figure}
    \centering
    \scalebox{0.6}{\begin{tikzpicture}
\begin{axis}[
    ylabel={Average Bandwidth Per Data Requester (Kbps)},
    xlabel={Replication Degree},
    xmin=0, xmax=14,
    ymin=0, ymax=80,
    xtick={0, 1, 2, 4, 8, 10, 12, 14},
    ytick={0, 10, 20, 30, 40, 50, 60, 70, 80},
    legend style={
    legend pos=outer north east, legend cell align = left},  
    ymajorgrids=true,
    grid style=dashed,
]

    \addplot[color = brown, mark size = 4, mark = star, mark color = red]
    coordinates 
    {
        (1,  1.21)
        (2,  2.70)
        (4,  8.31)
        (8,  29.63)
        (10, 46.48)
        (12, 60.42)
        (14, 76.58)
    };
    \addlegendentry{Pyramid} 
    
    \addplot[color = black, mark size = 4, mark = +, mark color = black]
    coordinates 
    {
        (1,  0.07)
        (2,  0.6)
        (4,  2.8)
        (8,  7)
        (10, 9.4)
        (12, 12.6)
        (14, 15.4)
    };
    \addlegendentry{Randomized}

    \addplot[color = blue, mark size = 3, mark = o, mark color = blue]
    coordinates 
    {
        (1,  0.84)
        (2,  1.96)
        (4,  4.2)
        (8,  20.4)
        (10, 30.8)
        (12, 39.6)
        (14, 48.2)
    };
    \addlegendentry{Cluster-based}

    \addplot[color = purple, mark size = 4, mark = +, mark color = purple]
    coordinates 
    {
        (1,  0.28)
        (2,  0.8)
        (4,  3.2)
        (8,  10.4)
        (10, 17.4)
        (12, 23.4)
        (14, 29.2)
    };
    \addlegendentry{Power-of-Choice}   
    
    \addplot[color = blue, mark size = 2, mark = square, mark color = orange]
    coordinates 
    {
        (1,  0.04)
        (2,  0.4)
        (4,  1.64)
        (8,  4.2)
        (10, 7)
        (12, 9.2)
        (14, 11.4)
    };
    \addlegendentry{GLARAS}   
    
    \addplot[color = green, mark size = 3, mark = square, mark color = blue]
    coordinates
    {
        (1,  0.92)
        (2,  2.08)
        (4,  7.14)
        (8,  26.2)
        (10, 38.22)
        (12, 48.6)
        (14, 65.66)
    };
    \addlegendentry{Correlation-based}


\end{axis}
\end{tikzpicture}}
    \caption{The performance of the replication algorithms with respect to the utility-awareness of replicas. The X-axis corresponds to the replication degree. The Y-axis represents the average available bandwidth of replicas per data requester at each time slot.}
    \label{pyramid:fig_utility}
\end{figure}

\subsection{Utility-Awareness}
As the utility-awareness metric, we measure the average available bandwidth per data requester, where the average is taken over replicas of all the data owners, during all the simulation time slots, and over the topologies under simulation. For a single data requester, the corresponding replica with respect to a data owner is the one with the minimum RTT to it among the replicas of the data owner. We compute the average available bandwidth for each data requester node by dividing its closest replica's bandwidth by the number of corresponding data requesters of that replica. Figure \ref{pyramid:fig_utility} represents the utility-awareness performance of the replication algorithms. 
\textit{For a specific replication algorithm, a point $(x,y)$ in Figure \ref{pyramid:fig_utility}
is interpreted as "placing $x$ replicas for a single data owner using this algorithm results on the average available bandwidth of $y$ Kbps per data requester node"}. As shown in the figure, by enforcing the replicas to be placed solely with respect to the latency distribution of the nodes in the underlying network, GLARAS performs even worse than the randomized distribution of the replicas in providing utility-awareness. Comparing the bandwidth of the two randomly chosen replication candidates, and replicating on the better one results in the power-of-choice strategy to outperform the randomized replication especially in the larger replication degrees. The performance gap among \mor, cluster-based and correlation-based, and the rest stems from the fact that these three replication algorithms have utility-based scoring. As the replication degree of the data owner increases, and the cluster-based replication divides the nodes into a larger number of cliques, the number of cliques with low availability also increases increases. This justifies the poor performance of the cluster-based replication compared to the correlation-based and \mor especially in higher replication degrees. Nodes in our simulations have the average online and offline duration of $2.7$ and $2.8$ hours, respectively. Mapping this to the uniform failure model results in the uniform failure probability of about $0.5$ for each node at each time slot. This results in the appearance of many anti-correlated pairs of nodes in the system, which are exploited by the correlation-based replication algorithm, and makes it as the best among the existing applicable decentralized solutions on the Skip Graph. \textbf{Compared to the best existing solutions, our proposed \mor improves the utility of replicas with a gain of about $\mathbf{\availabilityGain}$ times on average.}

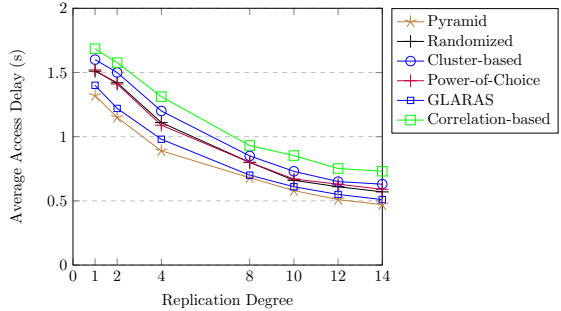
\begin{figure}
    \centering
    \scalebox{0.6}{\begin{tikzpicture}
\begin{axis}[
    ylabel={Average Access Delay (s)},
    xlabel={Replication Degree},
    xmin=0, xmax=14,
    ymin=0, ymax=2,
    xtick={0, 1, 2, 4, 8, 10, 12, 14},
    ytick={0, 0.5, 1, 1.5, 2},
    legend style={
    legend pos=outer north east, legend cell align = left},  
    ymajorgrids=true,
    grid style=dashed,
]

    \addplot[color = brown, mark size = 4, mark = star, mark color = red]
    coordinates 
    {
        (1,  1.32)
        (2,  1.15)
        (4,  .89)
        (8,  .68)
        (10, .58)
        (12, .51)
        (14, .47)
    };
    \addlegendentry{Pyramid} 
    
    \addplot[color = black, mark size = 4, mark = +, mark color = black]
    coordinates 
    {
        (1,  1.51)
        (2,  1.42)
        (4,  1.11)
        (8,  .8)
        (10, .66)
        (12, .61)
        (14, .57)
    };
    \addlegendentry{Randomized}

    \addplot[color = blue, mark size = 3, mark = o, mark color = blue]
    coordinates 
    {
        (1,  1.6)
        (2,  1.5)
        (4,  1.2)
        (8,  .85)
        (10, .73)
        (12, .65)
        (14, .63)
    };
    \addlegendentry{Cluster-based}

    \addplot[color = purple, mark size = 4, mark = +, mark color = purple]
    coordinates 
    {
        (1,  1.52)
        (2,  1.41)
        (4,  1.09)
        (8,  .8)
        (10, .67)
        (12, .63)
        (14, .59)
    };
    \addlegendentry{Power-of-Choice}   
    
    \addplot[color = blue, mark size = 2, mark = square, mark color = orange]
    coordinates 
    {
        (1,  1.4)
        (2,  1.22)
        (4,  .98)
        (8,  .7)
        (10, .61)
        (12, .55)
        (14, .51)
    };
    \addlegendentry{GLARAS}   
    
    \addplot[color = green, mark size = 3, mark = square, mark color = blue]
    coordinates
    {
        (1,  1.685)
        (2,  1.577)
        (4,  1.312)
        (8,  .931)
        (10, .853)
        (12, .752)
        (14, .731)
    };
    \addlegendentry{Correlation-based}


\end{axis}
\end{tikzpicture}}
    \caption{The performance of the replication algorithms with respect to the locality-awareness of replicas. The X-axis corresponds to the replication degree. The Y-axis shows the average latency between each data requester and its closest online replica at each time slot.}
    \label{pyramid:fig_locality}
\end{figure}

\subsection{Locality-Awareness}
Figure \ref{pyramid:fig_locality} shows the performance of replication algorithms with respect to the locality-awareness. At each time slot, we measure the locality-awareness as the average access latency between each data requester node and its closest online replica, which is performed for the replicas of every data owner at each time slot. We report the average access delay of replication per data owner at each time slot and over all the topologies under simulation. By purely considering the utility of the nodes, correlation-based replication that performs the second best in the utility-awareness, is the weakest one in the locality-awareness. As explained in Section \ref{pyramid:sec_simulation}, in our simulation setups, nodes with similar locality features show a similar availability behavior. Hence, by clustering the nodes into cliques based on their utility vectors -that also contains their availability behavior- and placing a replica in each clique, the cluster-based replication performs better than the correlation-based in locality-awareness. Cluster-based replication, however, is outperformed by the randomized and power-of-choice replications in locality-awareness. Since once the replica of a clique (temporarily) departs the system, the data requesters of that replica should benefit from the available replicas in the nearby cliques, which increases their average access delay. On the contrary, since randomized and power-of-choice replications choose the replicas uniformly with respect to their latency distribution, any subset of their available replicas follows a uniform distribution with respect to the locality. Among the existing solutions, by purely considering the locality-awareness of the replicas, GLARAS provides the minimum average access delay. GLARAS, however, distributes the replicas totally regardless of the utility behavior of the nodes. This results in time slots where crowded regions of the system remain without an available replica and are imposed to use other regions' replicas, which results in an increase on the average access delay of the replication. In contrast to GLARAS, \mor aims at minimizing the locality-awareness per time slot while considering the utility of replicas. \textbf{Compared to GLARAS that acts as the best existing locality-aware replication, our proposed \mor reduces the average access delay of replicas under churn with a gain of about $\mathbf{\localityGain}$ times on average.} Concerning the scalability, we also simulated \mor in system sizes of $1024$ and $2048$ nodes and observed consistent performance results with the $4096$ nodes setup in both the utility- and locality-awareness of replicas. We exclude the details of scalability results for the sake of space.

\subsection{Asymptotic Complexity of \mor}
\textbf{Memory Complexity:} Having the system capacity of $\systemCapacity$ nodes, the data owner that executes an instance of \mor only needs to maintain the the set of landmarks' features with the memory complexity of $O(\log{\systemCapacity})$, the replication degree with the memory complexity of $O(1)$, and the utility table of the system with memory complexity of $O(|\landmarkSet| \times \virtualSystemSize \times |UV|)$. Considering $\virtualSystemSize = \log{\systemCapacity}$ and $|UV|$ as a constant parameter of the protocol, the memory complexity of \mor is $O(\log^{2}{\systemCapacity})$.

\textbf{Time Complexity:} The asymptotic running time complexity of SWD is $O(\log^{2}{\systemCapacity})$ \cite{hassanzadeh2018decentralized}. TCWD function of \mor iterates twice over a table of size $\virtualSystemSize \times |UV|$, which causes a time complexity of $O(\log{\systemCapacity})$. The ILP part of \mor that is represented by Equations \ref{pyramid:objective}-\ref{pyramid:cons5} has both an objective function size and the number of constraints of $O(|UV| \times \virtualSystemSize^{2})$. Considering that $|UV|$ is a constant parameter, the ILP size of \mor is $O(\log^{2}{\systemCapacity})$ in terms of the size of the objective function and the number of constraints. Running on Intel i5 $2.60$ GHz CPU and $8$ GB of RAM and using lpsolve \cite{lpsolve} to solve the ILP model of \mor, a single execution of \mor in a $4096$ nodes system takes the average computation time of about $3$ minutes to determine the placement of $14$ replicas for a single data owner.

\textbf{Communication Complexity:} By the communication complexity, we mean the round complexity (i.e., the number of rounds of the protocol), the message complexity (i.e., the total number of the transmitted messages), and the bit complexity (i.e., the total number of transmitted bits), altogheter. In our system model, a single node updates the utility table of the system by reporting its utility vector on the underlying aggregation scheme (i.e., LightChain) once during every cycle of \fpti (i.e., once every $24$ hours based on our simulation setup), which causes the communication complexity of $O(\log{\systemCapacity})$ \cite{hassanzadeh2019lightchain}. As an input to \mor, a data owner requires retrieving the latest state of the utility table of the system, which causes a communication complexity of $O(\log{\systemCapacity})$ \cite{hassanzadeh2019lightchain}. Having the replication degree of $r$, \mor performs $r$ searches for utility to map the replicas from the virtual system to the original system. Each search causes a communication complexity of $O(\log{\systemCapacity})$ \cite{hassanzadeh2016laras}. Hence, placing $r$ replicas for a single data owner using \mor costs an $O(r \times \log{\systemCapacity})$ communication complexity. 

\section{Conclusion}
\label{pyramid:sec_conclusion}
To improve the response time of Skip Graph-based P2P cloud storage systems under churn, we proposed \mor, which is the first fully decentralized utility- and locality-aware replication algorithm for Skip Graph overlays. In replica placement, \mor considers the heterogeneity of the nodes with respect to their latency distribution in the underlying network, availability behavior, storage capacity, and bandwidth. \mor enables a data owner to place its replicas in a fully decentralized manner, and in a way that the \goal of replicas are achieved under churn. \mor is utility-aware as it maximizes the average available bandwidth of replicas per time slot.
Additionally, \mor is locality-aware as it minimizes the average response time of replicas.
Our simulation results show that compared to the best existing replication algorithms that are applicable on a Skip Graph-based P2P overlay, \mor improves the \goal of replicas with a gain of about \textbf{\availabilityGain and \localityGain times}, respectively. 
\section*{Acknowledgement}
The authors thank Y\"{u}\c{s}a \"{O}mer Altıntop, Berkay Barlas, and Esat Tunahan Tuna for their contributions to the SkipSim implementation.



\bibliographystyle{IEEEtran}
\bibliography{references}
\end{document}